\documentstyle[11pt]{article}
\def\1{\mbox{I\hspace{-.15em}1}}

\def\R{{\rm I\hspace{-.15em}R}}

\def\b{\begin{equation}}
\def\e{\end{equation}}
\def\bee{\begin{enumerate}}
\def\eee{\end{enumerate}}

\title{N=2 de Sitter Supersymmetry Algebra}
\author{O. Jalili$^{1}$\thanks{E-mail: omid\_jalili@yahoo.com}, S. Rouhani$^{1}$}
\date{\today}

\begin{document}
  \maketitle {\centerline{\it$^1 $ Plasma
Physics Research Center, Islamic Azad University,}
 \centerline{\it P.O.BOX 14835-157, Tehran,
IRAN} }
\begin{abstract}

It was shown that $N=1$ super-symmetry algebra can be constructed in
de Sitter space \cite{parota}, through calculation of charge
conjugation in the ambient space notation \cite{morota}. Calculation
of $N=2$ super-symmetry algebra constitutes the main frame of this
paper. $N=2$ super-symmetry algebra was presented in \cite{piniso}.
In this paper, we obtain an alternative $N=2$ super-symmetry
algebra.

\end{abstract}

\vspace{0.5cm} {\it Proposed PACS numbers}: 04.62.+v, 98.80.Cq,
12.10.Dm \vspace{0.5cm}

\section{Introduction}

Recent astrophysical data confirm that our universe could be
suitably represented by de Sitter universe. Extensive studies have
been dedicated to Q.F.T. in de Sitter universe. Pilch et al
concentrated on super-symmetry (susy) in de Sitter universe
\cite{piniso}. Drawing similarities with Minkowskian space they
selected a special charge conjugation, which enabled to closed the
supper algebra for even value of $N$. By symmetry consideration of
spinor field in de Sitter ambient space notation, a new charge
conjugation is obtained \cite{morota}. It is shown that $N=1$
super-symmetry algebra can be constructed in this case
\cite{parota}.

In the present work we have considered $N=2$ super-symmetry
algebra in de Sitter ambient space notation. After introducing the
ambient space notation (section 2), the general de Sitter
super-algebra is recalled in section $3$. In section $4$, we have
shown that in this case two different algebras do exist; $(1)$ one
is the same as previous super-symmetry algebra \cite{piniso} and
$(2)$ the other is an alternative super-symmetry algebra,
mentioned but not explored in \cite{piniso}. Finally, a brief
conclusion and an outlook have been discussed in section $5$.

\section{Notation}

The de Sitter space is an elementary solution of the positive
cosmological Einstein equation in the vacuum. It is conveniently
seen as a hyperboloid embedded in a five-dimensional Minkowski
space
     \b X_H=\{x \in \R^5 ;x^2=\eta_{\alpha\beta} x^\alpha x^\beta
=-H^{-2}\},\;\;
      \alpha,\beta=0,1,2,3,4, \e
where $\eta_{\alpha\beta}=$diag$(1,-1,-1,-1,-1)$. The de Sitter
metrics reads $$
ds^2=\eta_{\alpha\beta}dx^{\alpha}dx^{\beta}\mid_{x^2=-H^{-2}}=g_{\mu\nu}^{dS}dX^{\mu}dX^{\nu},\;\;
\mu=0,1,2,3,$$ where the $X^\mu$'s are the $4$ space-time
intrinsic coordinates on dS hyperboloid. Different coordinate
systems can be chosen for $X^\mu$. A $10$-parameter group
$SO_0(1,4)$ is the kinematical group of the de Sitter universe. In
the limit $H=0$, this reduces to the Poincar\'e group. There are
two Casimir operators
    $$ Q^{(1)}=-\frac{1}{2}M_{\alpha\beta} M^{\alpha\beta} ,$$
     \b
Q^{(2)}=-W_{\alpha}W^{\alpha},\;\;\;W_{\alpha}=-\frac{1}{8}\epsilon_
  {\alpha\beta\gamma\delta\eta}M^{\beta\gamma} M^{\delta\eta}, \e
where $\epsilon_ {\alpha\beta\gamma\delta\eta}$ is the usual
antisymmetrical tensor and the $M_{\alpha\beta}$'s are the
infinitesimal generators, which obey the commutation relations
\begin{equation} \lbrack M_{\alpha\beta},M_{\gamma\delta}\rbrack
= -i(\eta_{\alpha\gamma}M_{\beta\delta}+\eta_{\beta\delta}
M_{\alpha\gamma}-\eta_{\alpha\delta}M_{\beta\gamma}-\eta_{\beta\gamma}
M_{\alpha\delta}).\label{eqcommf}
\end{equation} $M_{\alpha\beta}$
can be represented as
$M_{\alpha\beta}=L_{\alpha\beta}+S_{\alpha\beta}$, where $
L_{\alpha\beta}=-i(x_{\alpha}\partial_{\beta}-x_{\beta}\partial_{\alpha})$
is the ``orbital" part and $S_{\alpha\beta}$ is  the ``spinorial"
part. The form of the $S_{\alpha\beta}$ depends on the spin of the
field. For spin $\frac{1}{2}$ field, it can be defined as
\begin{equation}
S_{\alpha\beta}=-{i\over
4}\lbrack\gamma_{\alpha},\gamma_{\beta}\rbrack, \label{genspin}
\end{equation}
where the five  $4\times 4$ matrices $\gamma^{\alpha}$ are the
generators of the Clifford algebra based on the metric
$\eta_{\alpha\beta}$: \begin{equation}
\gamma^{\alpha}\gamma^{\beta}+\gamma^{\beta}\gamma^{\alpha}=2\eta^{\alpha\beta}{\1}\,
,
\quad{\gamma^{\alpha}}^{\dag}=\gamma^{0}\gamma^{\alpha}\gamma^{0}.\label{Clifford}
\end{equation}
An explicit and convenient representation is provided by
\cite{ta1,brgamota,taka} $$ \gamma^0=\left( \begin{array}{clcr} \1
& \;\;0 \\ 0 &-\1 \\ \end{array} \right)
 ,\gamma^4=\left( \begin{array}{clcr} 0 & \1 \\ -\1 &0 \\ \end{array} \right),  $$
  \b   \gamma^1=\left( \begin{array}{clcr} 0 & i\sigma^1 \\ i\sigma^1 &0 \\    \end{array} \right)
     ,\gamma^2=\left( \begin{array}{clcr} 0 & -i\sigma^2 \\ -i\sigma^2 &0 \\  \end{array} \right)
      , \gamma^3=\left( \begin{array}{clcr} 0 & i\sigma^3 \\ i\sigma^3 &0 \\   \end{array} \right),\e
where $ \1 $ is the unit $ 2\times2 $ matrix and $\sigma^i $ are
the Pauli matrices. This representation had been proved to be
useful in analysis of the physical relevance of the group
representation \cite{brgamota}. In this representation,
$$\quad{\gamma^{\alpha}}^{T}=\gamma^{4}\gamma^{2}\gamma^{\alpha}\gamma^{2}\gamma^{4}.$$

The spinor wave equation in de Sitter space-time has been
originally deduced by Dirac in 1935 \cite{dir}, and can be
obtained from the eigenvalue equation of the second order Casimir
operator \cite{ta1,brgamota}
\begin{equation}  (-i\not x\gamma.\bar{\partial} +2i+\nu)\psi(x)=0,
\end{equation}
where $\not x=\eta_{\alpha \beta} \gamma^\alpha x^\beta$ and
$\bar{\partial_{\alpha}}=\partial_{\alpha}+H^2x_{\alpha}x\cdot\partial$.
Due to covariance of the de Sitter group, the adjoint spinor is
defined as follows \cite{brgamota}:
\begin{equation} {\overline \psi}(x)\equiv
\psi^{\dag}(x){\gamma^0}{\gamma^4}.\label{adj} \end{equation}

The charge conjugation spinor $\psi^c$ was calculated in the
ambient space notation \cite{morota} \b \psi^{c}=\eta_c
C(\gamma^4)^T(\bar \psi)^T,\e where $\eta_c$ is an arbitrary
unobservable phase value, generally set to unity. It is easily
shown that $\psi^c$ transforms in the same way as $\psi$
\cite{morota}
$$\psi'^c(x')=g(\Lambda)\psi^c(x).$$ The wave equation of $\psi^c$
is different from the wave equation of $\psi$ by the sign of the
''charge'' $q$ and $\nu$ \cite{morota}. Thus it follows that if
$\psi$ describes the motion of a dS-Dirac ''particle'' with the
charge $q$, $\psi^c$ represents the motion of a dS-Dirac
''anti-particle'' with the charge $(-q)$. In other words $\psi$
and ${\psi}^c$ can describe "particle'' and "antiparticle'' wave
functions. $\psi$ and ${\psi}^c$ are charge conjugation of each
other \b ({\psi}^c)^c=C\gamma^0{{\psi}^c}^\ast=C\gamma^0(C
\gamma^0\psi)=\psi. \e

In the present framework charge conjugation is an antilinear
transformation. In the $\gamma$ representation $(6)$ we had
obtained \cite{morota}:
$$ C\gamma^{0}C^{-1}=\gamma^0 , C\gamma^4C^{-1}=-\gamma^4$$ \b
C\gamma^{1}C^{-1}=\gamma^1,C\gamma^{3}C^{-1}=\gamma^3,C\gamma^{2}C^{-1}=-\gamma^2.
\e In this representation $C$ anticommutes with $\gamma^2$ and
$\gamma^4$ and commutes with other $\gamma$-matrix therefore the
simple choice may be taken to be as $C=\gamma^2\gamma^4$, where
the following relation is satisfied \b C=-C^{-1}=-C^T=-C^\dag.\e
This clearly illustrates the non-singularity of $C$.

\section{Super-symmetry Algebra}

Super-symmetry in de Sitter space has been studied by Pilch et al
\cite{piniso}. Recently, it has been investigated in constant
curvature space as well \cite{mcsh,fell}. In this section we have
presented the super-symmetry algebra in ambient space notation. It
is shown that if the spinor field and the charge conjugation
operators are defined in the ambient space notation, a new de
Sitter super-algebra can be attained.

In order to extend the de Sitter group, the generators of
super-symmetry transformation $Q^n_i$ are introduced. Here $i$ is
the spinor index ($i=1,2,3,4$) and $n$ is the super-symmetry index
$ n=1,..,N$. $Q^n_i$'s are super-algebra spinor generators which
transform as de Sitter group spinors. The generators
$\tilde{Q}^n_i$ are defined by \b \tilde{Q}_i=\left(Q^T \gamma^4
C\right)_i=\bar{Q^c}_i,\e where $Q^T$ is the transpose of $Q$. It
can be shown that $\tilde{Q}\gamma^4 Q$ is a scalar field under
the de Sitter transformation \cite{morota}. For $N\neq 1$, closure
of algebra requires extra bosonic generators. These do not
necessarily commute with other generators and consequently are not
central charges. They are internal symmetry generators
\cite{piniso}. These generators, shown by $T_{mn}$, commute with
de Sitter generators.

Therefore the de Sitter super-algebra in four-dimensional
space-time has the following generators:
\begin{itemize}
    \item$M_{\alpha\beta}$, the generators of de Sitter group,
    \item the internal group generators $T_{nm}$, that are defined by
    the additional condition
$$T_{nm}=-T_{mn}; n,m=1,2....N,$$
    \item the $4$-component dS-Dirac spinor generators,
$$Q^n_i,\;\;\;i=1,2,3,4,\;\;\; n=1,2,...,N.$$
 \end{itemize}

To every generator $A$, a grade $p_a$ is assigned. For the
fermionic generator $p_a=1$, and for the bosonic generator
$p_a=0$. The bilinear product operator is defined by \b
[A,B]=-(-1)^{p_a.p_b} [B,A].\e The generalized Jacobi identities
are \b (-1)^{p_a.p_c} [A,[B,C]]+(-1)^{p_c.p_b}[C,
[A,B]]+(-1)^{p_b.p_a} [B,[C,A]]=0.\e Using different generalized
Jacobi identities, similar to the method presented by Pilch el al.
\cite{piniso}, the full dS-superalgebra can be written in the
following form \cite{morota}:\b
 [M_{\alpha\beta}, M_{\gamma\delta}] =
-i(\eta_{\alpha\gamma}M_{\beta\delta}+\eta_{\beta\delta}
M_{\alpha\gamma}-\eta_{\alpha\delta}M_{\beta\gamma}-\eta_{\beta\gamma}
M_{\alpha\delta}),$$ $$
[T_{rl},T_{pm}]=-i(\omega_{rp}T_{lm}+\omega_{lm}T_{rp}-\omega_{rm}T_{lp}-\omega_{lp}T_{rm}),$$
$$ [M_{\alpha\beta}, T_{rl}] =0,$$
$$[Q^{r}_{i},M_{\alpha\beta}]=(S_{\alpha\beta}Q^{r})_{i},\;\;
[\tilde Q^{r}_{i},M_{\alpha\beta}]=-(\tilde Q^{r}
S_{\alpha\beta})_{i}$$
$${[Q^{r}_{i},T^{lp}]}={-(\omega^{rl}Q^{p}_{i}-\omega^{rp}Q^{l}_{i})}$$
$$ \{Q^{r}_{i},Q^{l}_{j}\}=\omega^{rl}\left(S^{\alpha\beta}\gamma^4
\gamma^2\right)_{ij}M_{\alpha\beta}+ \left(\gamma^4
\gamma^2\right)_{ij}T^{rl},\e where $S_{\alpha\beta}$ is a matrix,
eq. $(4)$, that satisfying the same algebra as $M_{\alpha\beta}$.
It is necessary to obtain matrix $\omega$, which determines the
structure of the internal group. In the next section, $N=2$
super-symmetry algebra is considered.

\section{$N=2$ super-symmetry algebra}

For $N=2$, the internal generators $T_{mn}$ are, $T_{11}=T_{22}=0$
and $T_{12}=-T_{21}=T$. There are two possibilities for $\omega $,
$
\mathrm{\omega}=\left( \begin{array}{clcr} 0 & 0 \\ 0 &0 \\
\end{array} \right),\;\; \mbox{and }\;\;
\mathrm{\omega}=\left( \begin{array}{clcr} 1& 0\\ 0 &1 \\
\end{array} \right).
$ For the trivial case $\omega=0$, the de Sitter super-symmetry
algebra, is determined by the following relations:
$$ [M_{\alpha\beta}, M_{\gamma\delta}] =
-i(\eta_{\alpha\gamma}M_{\beta\delta}+\eta_{\beta\delta}
M_{\alpha\gamma}-\eta_{\alpha\delta}M_{\beta\gamma}-\eta_{\beta\gamma}
M_{\alpha\delta}),$$ $$[T_{rl},T_{pm}]=0, \;\;\; [M_{\alpha\beta},
T_{rl}] =0 ,\;\;\;{[Q^{r}_{i},T^{lp}]}=0,$$ $$
[Q^{r}_{i},M_{\alpha\beta}]=(S_{\alpha\beta}Q^{r})_{i},\;\;
[\tilde Q^{r}_{i},M_{\alpha\beta}]=-(\tilde Q^{r}
S_{\alpha\beta})_{i}, $$ \b \{Q^{r}_{i},Q^{l}_{j}\}=\left(\gamma^4
\gamma^2\right)_{ij}T^{rl}. \e This is proven by the use of
generalized Jacobi identities. This is indeed a new super-algebra.

For $\omega=\1$, the de Sitter super-symmetry algebra is \b
 [M_{\alpha\beta}, M_{\gamma\delta}] =
-i(\eta_{\alpha\gamma}M_{\beta\delta}+\eta_{\beta\delta}
M_{\alpha\gamma}-\eta_{\alpha\delta}M_{\beta\gamma}-\eta_{\beta\gamma}
M_{\alpha\delta}),$$ $$
[T_{rl},T_{pm}]=-i(\delta_{rp}T_{lm}+\delta_{lm}T_{rp}-\delta_{rm}T_{lp}-\delta_{lp}T_{rm}),$$
$$[Q^{r}_{i},M_{\alpha\beta}]=(S_{\alpha\beta}Q^{r})_{i},\;\;
[\tilde Q^{r}_{i},M_{\alpha\beta}]=-(\tilde Q^{r}
S_{\alpha\beta})_{i}$$
$${[Q^{r}_{i},T^{lp}]}={-(\delta^{rl}Q^{p}_{i}-\delta^{rp}Q^{l}_{i})},\;\;\;[M_{\alpha\beta}, T_{rl}] =0,$$
$$ \{Q^{r}_{i},Q^{l}_{j}\}=\delta^{rl}\left(S^{\alpha\beta}\gamma^4
\gamma^2\right)_{ij}M_{\alpha\beta}+ \left(\gamma^4
\gamma^2\right)_{ij}T^{rl}.\e It is interesting to note that if we
multiply these relations with $\gamma^4$, we obtained exactly the
Pilch's result \cite{piniso}. This is due to the fact that adjoint
spinors are defined in two different definition ways: $$
{\overline \psi}(x)\equiv \psi^{\dag}(x){\gamma^0}{\gamma^4},\;\;
\mbox{our definition}, $$
$$ {\overline \psi}(x)\equiv
\psi^{\dag}(x){\gamma^0}, \;\;\mbox{usual definition}.$$ This
choice of adjoint spinor has been motivated by the covariance of
de Sitter group, contrary to the usual mikowskian analogy
\cite{brgamota}.

\section{ Conclusions}

A new charge conjugation has been obtained in the quantization
process of the spinor field in de Sitter space constructed through
ambient space notation \cite{morota,brgamota}. It is shown that an
alternative $N=2$ super-symmetry algebra, mentioned but not explored
in \cite{piniso}, can be attained by the use of the spinor fields
and charge conjugation in the ambient space notation. This result
can be used for construction the simplest $N=2$ de Sitter
supergravity lagrangian, which will be considered in the forthcoming
paper.

\vspace{0.5cm} \noindent {\bf{Acknowledgements}}: Authors would
like to convey their gratitude to Dr. Mahmoudi and Research Center
of Azad University at Noor.
 
\end{document}